# ERP project's Internal Stakeholder network and how it influences the project's outcome

Kristian Jääskeläinen, Accenture Finland; and L-F Pau, Prof. Mobile business, Copenhagen Business School and Rotterdam school of management lfp@cbs.dk

**ABSTRACT**

So far little effort has been put into researching the importance of internal ERP project stakeholders' mutual interactions, realizing the project's complexity, influence on the whole organization, and high risk for a useful final outcome. This research analyzes the stakeholders' interactions and positions in the project network, their criticality, potential bottlenecks and conflicts. The main methods used are Social Network Analysis, and the elicitation of drivers for the individual players. Information was collected from several stakeholders from three large ERP projects all in global companies headquartered in Finland, together with representatives from two different ERP vendors, and with two experienced ERP consultants. The analysis gives quantitative as well as qualitative characterization of stakeholder criticality (mostly the Project Manager(s), the Business Owner(s) and the Process Owner(s)) , degree of centrality, closeness , mediating or bottleneck roles, relational ties and conflicts (individual, besides those between business and project organizations) , and clique formations. A generic internal stakeholder network model is established as well as the criticality of the project phases. The results are summarized in the form of a list of recommendations for future ERP projects to address the internal stakeholder impacts .Project management should utilize the latest technology to provide tools to increase the interaction between the stakeholders and to monitor the strength of these relations. Social network analysis tools could be used in the projects to visualize the stakeholder relations in order to better understand the possible risks related to the relations (or lack of them).

**INTRODUCTION**

Several organizations are facing demanding company-wide Enterprise resource planning (ERP) projects in the coming years. ERP projects are expensive, complex and influence the whole organization. The challenge is to make them successful. There are many examples of failed attempts which have cost millions of dollars without bringing the benefits they were supposed to.

There has been a lot of research on ERP project's critical success factors and the results have emphasized the importance of top management support, ERP systems architecture flexibility, effective communication and change management, organization participation, minimal customization just to name a few. Nah et al. (2001) and others have looked at several ERP projects and concluded that there are 11 factors that are critical to ERP project implementation; most of those listed are technical or relate to the overall information management strategy of the company, procedures, policies and standards .

So far little effort has been put in researching the importance of the project stakeholders individually or as a network (Rowley, 1997) (ITToolbox, 2006). In an ERP project, the social relations between the stakeholders become essential: stakeholders form a network consisting of relations; influencing each other, the decisions made during the project and the final outcome. Each stakeholder has his or her own drivers, based on which they act during the project.

Potential drivers, which influence the decision making of an individual stakeholder, could be fear of loosing power in the organization but also be the opposite, will to increase own power, which could be achieved for instance by performing well in the project and gaining recognition from other stakeholders. Other potential drivers could be the fear of being held responsible of a decision, influencing the willingness to make decisions, and on the other hand willingness to influence areas where the stakeholder doesn't have the end responsibility. In sum, each individual has his own drivers and objectives trying to influence the other stakeholders so that those objectives would be realized.

Although having many commonalities with other IS projects, ERP projects differ in many ways from other projects by being so comprehensive from the organizational point of view. An ERP project normally involves company-wide business process redesign and covers most of the company's processes influencing almost everyone in the organization.

So far much of the research on stakeholder issues in ERP projects has been drifting too far from current issues. By current issues are meant the problems, which the IS professionals face in their daily work. This research is aiming at providing concrete conclusions and recommendations for the ERP project professionals in order to avoid, or at least to mitigate, the project's risks related to the stakeholders and their relations.

The paper addresses as main research question:
- What is the ERP project's Internal Stakeholder network and how does it influence the project's outcome?

Sub questions, which further define the main question, are:

- Who are the key stakeholders inside an ERP project once launched?
- Who are the most critical stakeholders inside an ERP project?
- Who are the potential intra-personal conflicts and bottlenecks that can affect the project?
- What are risks related to the internal stakeholders and their position in the network?
- What are the drivers behind internal stakeholders' actions?
- What type of relations exists between the stakeholders?

This experimental research focuses on three large company-wide ERP projects at companies headquartered in Finland, which have all already implemented ERP systems ranging from thousand to several thousands of ERP users. All of the projects cover the main business processes, such as finance and controlling, demand planning and manufacturing, sales and logistics etc.

The originality of this research is based on the fact that it is using a theoretical approach which has not been used widely in information systems research, since it belongs traditionally to sociological sciences, namely Social network analysis (SNA). However, in sociology as well as in communications traffic analysis this theory is widely used in order to research human relations. An ERP project if any is based on human relations so therefore the Social Network Analysis theory is seen as relevant. SNA has the ability to describe the <u>real stakeholder relations</u> inside a project instead of the traditional approach which is based mainly on the official organization structures on one hand and the contractual aspects on the other hand. When only looking at the official organization and project execution structures many relevant details of the stakeholders' interaction with each other are ignored.

Supported by the SNA theory, incentive analysis, and metrics collected during the analysis of three large ERP projects in Finland, this research explains the ERP project's key stakeholders, their motives, relations between the stakeholders; identifies the most critical roles, possible conflicts and bottlenecks in the project organization. This information helps to better understand how the project's internal stakeholder relations influence an ERP project and how the risks (time, costs, quality of the ERP implementation, error rates, etc ...) related to the stakeholder relations could be minimized.

The three companies which have been investigated differ heavily from each other. The first one is a well known large services company acting only in the Finnish markets, the second is a middle sized traditional manufacturing company with international focus, the third one a large international hi-tech company. The differences between the companies have allowed gathering specific detailed information from three different organizational environments.

The research population is the internal ERP project stakeholders in the three different projects (one project per company), completed with stakeholders external to the three deployment companies for validation of the results, including representatives from two ERP-software providers and from a systems integration consultancy company.

After a survey and discussion of the relevance of this research, the paper presents the methodology used with the cases and the information collection. The analysis provides quantitative and qualitative results, leading to a list of recommendations for future ERP projects, before some conclusions.

**SURVEY AND RELEVANCE OF THE RESEARCH**

Enterprise Resource Planning (ERP) has been a hot topic since the early 90's. Since many implementations, costing millions of dollars have failed to bring the intended benefits there is a need to conduct research on which factors influence the implementation process. ERP projects are a complex mixture of technology, business, organization and politics. Therefore there are many factors that influence their success or failure. In this Section the relevance of the research problem is explained by demonstrating the economical scale and importance of the ERP projects and how big implications a project failure might have for a company.

An average ERP project lasts normally between one and three years (Komiega, 2001; Darwin's Executive Guide, 2004). It is difficult to say accurately how much ERP projects cost in average, due to the fact that the projects differ a lot from each other for instance scope wise. Nevertheless, there has been research indicating the total costs of ownership for an ERP project. Meta Group made a research surveying 63 companies from small to large in different industries and their conclusion was that an average project's total costs, from the beginning of the project till two years after the project completion, was US$ 15 Million, the

average cost per user being as high as $53 000 (Darwin's Executive Guide, 2004).

The rate of success of the projects has been researched by many. According to Rao (2000) as many as 96,4 percent of ERP implementations fail. Supporting the high failure rate, but with a different rate, Al-Mashari (2000) concluded that 70 percent of ERP implementations fail to achieve the estimated benefits. According to Koch (2002) 40 percent of ERP project managers failed to achieve their original business case even after the ERP system being live for a year or more. Over 20 percent stop their projects before completion. Even in the projects who claim being successful, costs were on average 25 % over budget and annual support costs went up by an average of 20 % over the legacy systems they replaced. A survey by Robbins-Gioia consultancy concluded that 51 % of companies were not satisfied with the ERP project results. Although many reports indicate that ERP projects fail easily not everyone agrees with this view amongst vendors, consultants but also users. AMR Research Inc. analyst Jim Shepherd claims that the end result is positive in most cases in the long run. According to him in almost every case, when you encounter a story about a failed ERP project, if you went back a year later, you would find that they are happily using the system (Robb, 2006).

Whether the high failure rate is accurate or not there are plenty of examples of major problems when implementing a new ERP system even when learning is present. Probably one known failure is the US fourth largest distributor of pharmaceuticals FoxMeyer Drugs' SAP R/3 project followed by a bankruptcy in 1996 due to major issues with e.g. inventory management (Davenport, 1998). Another example of an unsuccessful ERP project is from the Norwegian Defense Forces. Their spokesman Sigurd Frisvold said in January 2005 that the ERP project exceeded its budget only in 2004 alone by approximately €100 Million. The Defense Forces admitted that the reason for the problems was neither the platform nor the software but the failure to adjust the organization to the new system (Savolainen, 2005). An example from Finland is KCI Konecranes who had a € 50 Million dispute with Baan after a failed Omniman ERP-system implementation in 2000. After a long legal fight in several countries the companies were able to reach an agreement. The result of the agreement was kept secret (Torikka, 2003).

ERP project issues might have dramatic consequences for a company. Hershey

Foods, one of the largest candy companies in US found out about this in the worst possible way. Hershey's new SAP system, Siebel CRM system and Manugistics supply chain software project, costing $112 Million, failed causing problems for the whole order and distribution system and serious problems for the business. The company faced a situation around Halloween where it could loose orders totaling $100 Million. When Wall Street heard of the problems Hershey's stock instantly dove with 8% in one day. Eventually it was found out that Hershey's problems were not unique, but the same occurred in most projects. Hershey just happened to have a bad timing for the problems just before high sales Halloween. The lesson Hershey management learned was that the system implementation is "easy". The difficult part is to get the personnel to change the way they are working. But eventually they will adapt. Other lesson was that ERP software is not just software. ERP changes the way the company conducts the business. (Koch, 2002)

What must be realized when reading these "horror stories" is that some business areas might have more difficulties in implementing ERP systems than others due to more complex or unexpected turns in the business environment. There is a difference between implementing an ERP system in a traditional manufacturing company with standard processes, which are "easier to control" than in a company selling services to consumers in a fast changing business environment. That is why there are less stories about major challenges for instance from the oil or metal industries than in health care, telecommunications services or consumer goods industries.

The experiences of failing can also derive from wrong expectations. Companies do not necessarily understand what to expect from the new system. Quite surprisingly a majority of the companies do not put much effort in calculating the breakdown of benefits the new system is supposed to bring. Bradford and Richtermeyer (2002) found out in their research that 57% of companies who invest in an ERP-system do not make detailed calculations of the benefits, in other words the business case is not made thoroughly.

What is common to the failure stories is that all of them emphasize the importance of the organizational aspect. None of the projects blame only the ERP software, hardware or the vendor who is implementing the system. In most cases not enough attention has been paid to the organizational aspect, how to make sure that the organization is ready for the change. Even though the most critical success factors have been listed by many researchers, many have pointed out that more studies should be made about how internal power structures and networks influence Information System projects (Butcher and Clark 1999; Dhillon, 2003; Silver *et al.*, 1995). "What are the drivers behind stakeholders' decisions and actions?" and "How do the stakeholders influence each other?" are very relevant questions for all ERP projects.

Looking at a concrete example of stakeholders' importance, Nestlé USA faced severe problems with their ERP implementation in 1999 because the project management forgot to involve the stakeholders in the project. None of the groups that were going to be directly affected by the new processes and systems were represented in the key stakeholder team. This lead to a situation where the key stakeholders, from executives to factory floor workers, didn't know how to use the new ERP system and they didn't understand the new business processes either. The conflict escalated to what was the only way to solve it, which was to invite all the stakeholders together, discuss the problems thoroughly, and redesign the ERP solution (Worthen, 2002).

"Organization charts prescribe that work and information flow in a hierarchy, but network mapping reveals actually this flow through a vast web of informal channels." (Krebs, 2006). Rice and Aydin (1991) have looked at how much employees influence each others opinions about a new IS system. They concluded that attitudes towards an information system are socially influenced by the people to whom the employee has close relations. Even though these studies do provide useful information for the project management the subject should be further studied and that is what this research is aiming at doing. Even though most research, which measured the influence of a new Information System on an organization, have concluded that the stakeholders are in key position for the success, no one has tried in the context of an ERP implementations to analyze the stakeholder relations and their importance in more detail during the project execution; this would involve mapping the entire stakeholder network, the actors, the influence of relations on the actors, strength of these relations etc. This paper is using Social Network Analysis (SNA) theory to determine all these factors and to create an overall generic model, which can benefit the planning of future ERP projects.

Most of the early research in the area has concentrated on organization's resistance to change. Some have looked at the subject from a power balance perspective, looking at how a new Information System stirs up the current power balance in the organization. New information system's influence on the organization, its power structures and resistance to change in IS projects, have been researched amongst others by Keen (1981), Markus (1983), Markus and Pfeffer (1983) and Newman and Rosenberg (1985). Von Hellens, Nielsen and Beekhuyzen (2005) conducted a qualitative case study on power and politics in an ERP implementation. Their conclusion was that the new system influences the internal power balance between the stakeholders, which can have significant influences on the organization. Burkhardt and Brass (1990) and Rice and Aydin (1991) have used SNA in researching how new IS systems influence the organization and the internal power structure. According to Burkhardt and Brass (1990) a new IS system increases uncertainty. At the same time those who are able to mitigate the uncertainty gain more power. The increased uncertainty also increases the need to communicate about that uncertainty, which changes the social communication network. What this has to do with an ERP project is that ERP normally comes together with a business process redesign, which stirs the organization structure, in other words the internal power structure. Because of the business process redesign the power of certain employees or departments might diminish. Keen (1981), Markus (1983) demonstrated how an IS project influences the power structures by redistributing data (information). According to Hellens et al. (2005) a new IS can change also other aspects of an organization, such as communication paths, influence, and control. Similar effects exist inside an ERP project organization within the company, as obviously successful implementations raise the credibility and image of the project leaders inside the company or for new roles.

The research on ERP project's influence on power and politics does have much in common with the subject of this paper. However, the main difference between the previous research and this paper is the perspective. The previous research has concentrated on finding out how the company's power structure has changed because of the ERP project and what causes the resistance to change. This paper is concentrating on analyzing how the ERP implementation project's stakeholders and their mutual relations influence the ERP project internally once launched and its outcome. In other words this research is providing tools for ERP project management while the other research focuses on how the new ERP system influences the organization prior to and after a decision to launch an ERP project in one or several business divisions.

## CASES AND RESEARCH METHOD

*Method*

The research model's objective is to describe an ERP project's key internal stakeholders and their relations. Although the research is focusing on the internal stakeholders, the model includes also the stakeholders from the Business and Company organizations who are assumed to have a key role in an ERP project. The model's internal stakeholders are defined as the key project organization members and those business organization and company management members who are involved almost full time with the project (Figure 1).Are also considered as internal stakeholders Key Users which are specialists from the line user organizations with process knowledge, and who are assigned almost full time to the project. The circles with the stakeholder name in the middle indicate the key stakeholders, the lines their relations and the arrow heads of the lines the direction of communication and influence. The research aims at validating the model, the key internal stakeholders, their relations and the direction and the strength of the relation.

The motivators or drivers (circle around the stakeholders in the graph in Figure 1) demonstrate that each stakeholder's behavior is influenced by various factors, both personal and organizational. The drivers have three dimensions. The positive drivers make the stakeholders to support the project and the negative drivers create resistance. The third dimension is time. Timing of the project can influence the other two dimensions, since the project's timing might not fit in the overall plans of the stakeholders. Stakeholders' relative attitudes and behavior can be time dependent. At the bottom of the graph are the ERP project phases which are influenced by the internal stakeholder network.

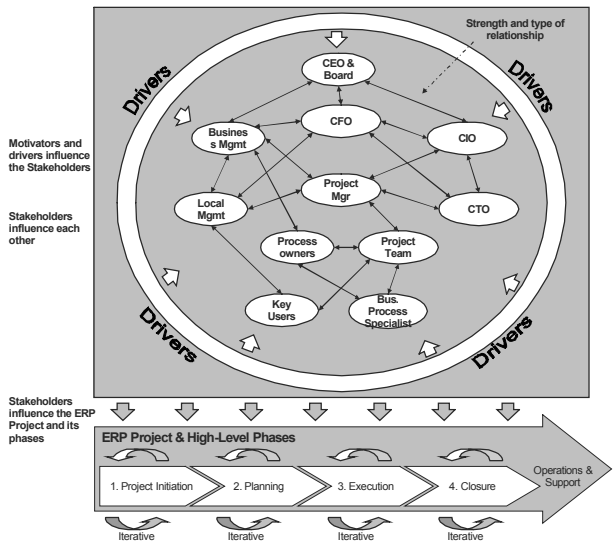

Figure 1: Internal stakeholder's network, stakeholders from the Business and Company organizations, and ERP lifecycle

*Cases*

The research focuses on three large company-wide ERP projects at three companies with headquarters in Finland, which have already implemented ERP systems ranging from thousand to several thousands of ERP users; all interviewees and most companies required to stay anonymous.

Project 1 was in a state owned company which sells services to consumers. The company has some manufacturing operations to produce the machines which are used to sell the services. There are approximately 1500 employees in the company. The market where the company is operating is fairly stable, in other words no major changes happen in its business environment. Prior to the unifying project start, there were more than 50 different ERP systems or procedures in use.

Project 2 is in a large international process-industry company which operates in more than 30 countries and has nearly 10 000 employees. The company transformed its business from local to the international markets in the past decade. In the recent years the company has focused on moving away from mass-produced towards value-added products and services to increase its competitiveness. Increasing globalization requires more efficient and flexible processes, which is one of the main reasons for the ERP project.

Project no 3 is in a large international business-to-business company manufacturing high-tech products. The company has grown fast in the past decade and the expectations for future growth are very positive, although the market is very competitive. The number of employees in the company is over ten thousand and the company operates in more than 50 countries. To face the future growth the management saw that there was a need for company-wide unified processes supported by an end-to-end ERP solution.

All of the projects cover the main processes, such as finance and controlling, demand planning and manufacturing, sales and logistics etc. In all of the projects the chosen ERP solution is SAP. However the research is not aiming at developing a model for only SAP projects but the purpose is to develop a general model, which can be utilized regardless of the chosen ERP software.

The companies in which the systems have been implemented differ heavily from each other. The differences between the companies are seen as strength for the research rather than a weakness, since this enables to gather relational and communications data from three different organizational environments. This should improve the validity of the results making the end results applicable for various ERP projects.

The research population is made of the key internal ERP project stakeholders in the three different projects, completed with stakeholders external to the three deployment companies, including representatives from two ERP-software providers (IFS Finland and SAP Finland) and from a systems integration consultancy company (Accenture Finland). It was not seen necessary to interview all the stakeholders that were involved in the project. It is believed that by interviewing initially identified key stakeholders a fairly reliable model of the stakeholders and their relations can be built. This model was then validated by the external stakeholders mentioned above. The objective was to reach a high level of external validity so that the model can be utilized in future ERP projects regardless of the organization. See Figure 2 below for the entire cycle from interviews till the creation of the generic model, and Figure 3 to get details of the interviewed internal stakeholders' roles in each case.

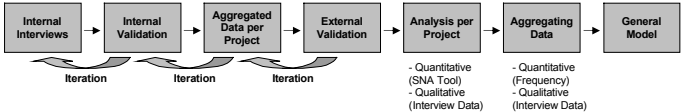

Figure 2: Experimental research methodology

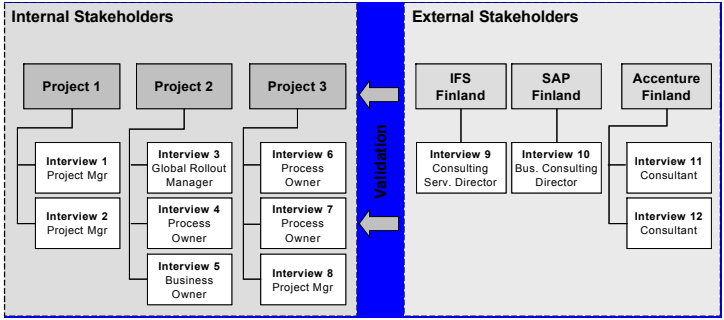

Figure 3: Internal and external ERP stakeholder roles and distribution of interviewees across projects

The boundaries of the research population are tied to the ERP project boundaries. This means that it only consists of persons who have been deeply involved with the project, in most cases full-time project members. As described above to map out relations, the identified key stakeholders are asked to further suggest who else should be interviewed. According to Laumann, Marsden and Prensky (1999) this is called a "*realist approach*". A realist approach focuses on actors and boundaries as perceived by the actors themselves.

*Information collection*

The main objective is to get a description of all the relations each interviewee had with other stakeholders. However, the interviewee is not only asked to describe his or her own contacts but to also identify relations between other stakeholders that he/she was aware of. The interviewee is then asked to further describe the nature of the relationship, how frequent the interaction was, what kind of interaction it was, who reported to whom, from whom did he/she get input to his work, to whom did he/she give input, was the relationship equal or did either party have more influence in the relationship, which were the most critical relationships from his/her point of view, which relationships lead to conflicts, and finally if the interviewee had recommendations on how to improve the project organization. Also any other comments the interviewee had were recorded. To improve the validity of the results each interviewee who has been involved in a given project is asked to comment on the results from the others preserving anonymity. Figure 4 shows how a graph from an interview looks like; it is drawn in Microsoft Visio. The "circles" indicate the stakeholders, the lines between the circles the relations, the arrows the direction of the relationship, the texts the intensity or frequency of the relations, the circles in grey the most critical stakeholders the interviewee identified, and the red lines the relations with conflicts. The interviewee is also asked to describe the strength of influence and the direction of the relation for those relations where the interviewee is able to give these values. The strength of influence is measured on a scale from 0 to 3, 0 being no influence, 1 being some influence, 2 strong influence and 3 very strong influence.

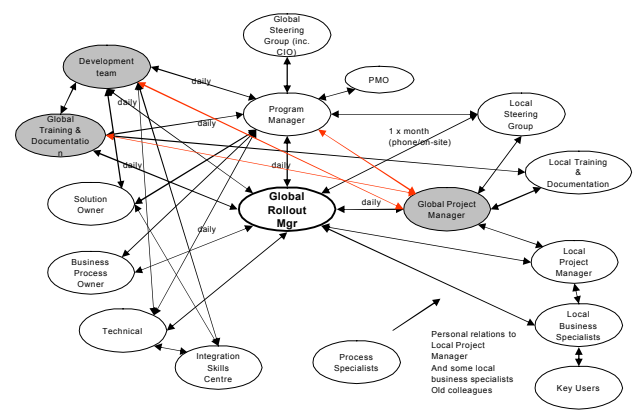

Figure 4: Example of an internal stakeholder interview graph

To analyze the project's social network the relation graph drawn in the interview must be transformed into a form suitable for analysis. A sociomatrix (Wasserman and Faust, 1994) is the data structure used to analyze the SNA data. A sociomatrix is created per project based on the graphs drawn like Figure 4 (which exist per project). The sociomatrix data, with the relation strength data, are entered into a commonly used Social Network Analysis tool, UCINET (Analytichtech, 2006). The UCINET data file can be entered into a SNA software tool called NetDraw (Analytichtech, 2006) which draws the stakeholder network, including the stakeholders, their relations and the strength of the relations. UCINET is also used to analyze and filter the data based on different predefined SNA formulas. After all the projects have been analyzed and the patterns have been identified, a generic model of internal ERP stakeholder relations, combining data from all three projects, has been created with simple combination majority logic.

## ANALYSIS

### Stakeholder Criticality

This Section is analyzing the stakeholder criticality combining two different approaches. The first approach is to use purely quantitative SNA methods to evaluate the criticality; the second approach is to analyze the interview data with a qualitative method involving having each interviewee state other stakeholder's criticality. The results of the two approaches are compared against each other.

The Degree of Centrality gives the number of links each node has, in other words how many relations each internal stakeholder has (Answers, 2007). Normally the more connections a stakeholder has the more important he is; but what matters rather is with which other stakeholders this stakeholder has relations which are the most important! Table 1 shows the Degree of Centrality per stakeholder role for all the three projects. The first column shows the project stakeholder role, the second Degree-column the absolute number of relations the stakeholder has, and the third column the Relative Degree of Centrality value in %, which is the ratio of the absolute Degree of Centrality to the maximum possible number of relations which the stakeholder could have (ego density), excluding the stakeholder himself.

| Project 1 | | |
|---|---|---|
| Stakeholder | Degree | Relative Degree |
| Project Manager | 10 | 83.333 |
| Program Steering Group | 6 | 50.000 |
| Vice President | 5 | 41.667 |
| Business Organization Mgr | 5 | 41.667 |
| Process Owner | 5 | 41.667 |
| Program Manager | 4 | 33.333 |
| Board | 3 | 25.000 |
| Development Team | 3 | 25.000 |
| Technical Team | 3 | 25.000 |
| Integration Team | 3 | 25.000 |
| Local Specialists | 2 | 16.667 |
| Key Users | 2 | 16.667 |
| Users | 1 | 8.333 |

| Project 2 | | |
|---|---|---|
| Stakeholder | Degree | Relative Degree |
| Business Process Owner | 15 | 71.429 |
| Program Mgr | 10 | 47.619 |
| Global Rollout Mgr | 10 | 47.619 |
| Global Project Mgr | 9 | 42.857 |
| Development Team | 9 | 42.857 |
| Technical Mgr | 7 | 33.333 |
| Local Steering Group | 5 | 23.810 |
| Global Training & Docum. | 5 | 23.810 |
| Local Project Mgr | 5 | 23.810 |
| Solution Owner | 5 | 23.810 |
| Local Business Specialists | 5 | 23.810 |
| Local Training & Docum. | 4 | 19.048 |
| Integration Team | 4 | 19.048 |
| Global Steering Group | 3 | 14.286 |
| Financial Services | 2 | 9.524 |
| Controller Community | 2 | 9.524 |
| Group Controller | 2 | 9.524 |
| Key Users | 2 | 9.524 |
| Financial Mgmt | 1 | 4.762 |
| PMO | 1 | 4.762 |
| FiCo Owner Group | 1 | 4.762 |
| IFRS Team | 1 | 4.762 |

| Project 3 | | |
|---|---|---|
| Stakeholders | Degree | Relative Degree |
| Solution_Owner | 10 | 45,455 |
| Business_Deployment_Mgr | 10 | 45,455 |
| Project_Manager | 10 | 45,455 |
| Process_Owner | 9 | 40,909 |
| Business_Owner_1 | 8 | 36,364 |
| Program_Manager | 8 | 36,364 |
| Sub-process_Owner | 7 | 31,818 |
| Integration_Manager | 6 | 27,273 |
| Business_Owner_2 | 6 | 27,273 |
| Other_Line_Org._Stakeholders | 6 | 27,273 |
| Vice_President | 6 | 27,273 |
| Program_Steering_Group | 5 | 22,727 |
| Regional_Solution_Owner | 4 | 18,182 |
| Country_Manager | 4 | 18,182 |
| CIO | 4 | 18,182 |
| Business_Owner_3 | 4 | 18,182 |
| Key_Users | 3 | 13,636 |
| Business_Owner_4 | 3 | 13,636 |
| CFO | 3 | 13,636 |
| Fi_&_Co_Project_manager | 3 | 13,636 |
| Configuration_Team | 3 | 13,636 |
| Program_Mgmt_Office | 2 | 9,091 |
| Operative_Board | 2 | 9,091 |

Table 1: Internal ERP Stakeholder centrality degree (absolute and relative)

In Project 1 the Project Manager stands out from the rest of the stakeholders. The Project Manager has a total of ten relations out of 12 possible relations, giving Relative Degree of Centrality of 83,3 %, that is a very central position in the project. The following two stakeholders are the Program Steering Group with 6 relations, having 50% of possible relations, followed by the Vice President with 5 relations and 41,7% of possible relations. As expected the Key Users, Local Specialist and Users have the least connections in the project, influencing their importance for the overall program. In Project 2 the Business Process Owner has the most relations, which is explained partly by the fact that the interview data included Finance and Controlling Process stakeholders who influence the Business Process Owner's decision making. The Degree of Centrality of the Program Manager, Global Rollout Manager, Global Project Manager, Development Team and of the Technical Manager are also clearly visible from the results. In Project 3 the differences between the stakeholders are not as significant as in the other two projects; it was clearly the largest project of the ones researched so it is natural that there are no stakeholders that would have relations to almost all other stakeholders. Solution owner, Project Manager and the Deployment Manager had all 10 relations, having 45,5% of all possible relations. The following most central stakeholders were the Process Owner with 9 relations, Program Manager, and Business Owner 1 with 8 relations, and the Sub-Process owner with 7 relations.

When comparing the projects against each other it can be noticed that in all of the projects the Project Managers stand high on the list, indicating their importance to the projects. However, when further analyzing the results it can be seen that there are clear differences between them.
There can be several reasons for the differences. One explanation could be that the project sizes are different; Project 1 is significantly smaller than Project 3 so it is easier for the Project Manager to have contact with almost everyone making the project organization relatively flat, while the Project Manager in Project 3 simply cannot have relations with all. Another explanation could be that the company cultures differ, so that in Company 1 the culture is more team-work based, while in Company 3 the culture is more top-down-based, meaning that the management gives instructions which go down through different hierarchical levels. The risk of having a "loose" organization is that the communication depends on the stakeholders who have a central role in the communication flow about the project.

In connection with the qualitative analysis, the SNA-data were compared to the data from the interviews. During the interviews each interviewee was requested to list the stakeholders that were the most critical for the project; aggregate lists were built and analyzed. Most of the interviewees couldn't see the criticality of the stakeholders beyond their own relations so they stated the stakeholders that were the most critical from their personal point of view. This can mean that they have missed some very critical stakeholders due to the fact that they did not know exactly what other stakeholders' role in the project was.

The Business Owner role was rated as critical by 9 of interviewees, Project Manager by 7, followed by the Development Team. Similarly to the Program Manager the sponsors, company management, were identified critical by the external interviewees (Figure 3). Even though the other interviewees did not mention the company management as being critical to the project execution, no one disagreed with this statement. The last two who got 2 and 1 votes were the Training Team and the Key Users.

*Closeness Centrality*

Closeness Centrality measures how "close" the stakeholder is to the other stakeholders in the network (Wasserman and Faust, 1994). Closeness Centrality is calculated by adding up all the distances a stakeholder has to other stakeholders; a Relative Closeness Centrality value in % can be determined by dividing previous value by total number of stakeholders minus one. By distance is meant how many steps (through other stakeholders) there are from the stakeholder to the other stakeholder in the network (Figure 4). The results of Closeness Centrality analysis support the previous conclusions about the differences between the projects. The higher the Relative Closeness Centrality value is, the "closer" the stakeholder is to the other stakeholders in the project organization. The mean Relative Closeness Centrality for Project 1 is 55,9%: the Project Manager has the highest Relative Closeness Centrality of 85,7% and the Key Users the lowest value of 35,3%. For Project 2 the mean Relative Closeness Centrality is 52,3%, the Business Process Owner having a value of 77,8% and the Key Users the lowest value of 38,9%. The Business Process Owner's value is again relatively high mainly because of the inclusion of the business process stakeholders in the data. When looking at the values for Project 3, the mean value is 44,8% , the highest 62,9% (Project Manager) and the lowest belongs to the Operative Board with a value of 37,3%. The low value of the Company Board is not surprising since they are not actively involved in the project but get regular status updates from the Vice President who is the only direct contact they have to the project. The conclusion is that the stakeholders in Projects 1 and 2 are significantly "closer" to each other than in Project 3. Again this could be explained by the difference in the project sizes, but when taking the interview data from Project 3's stakeholders into account, the conclusion that the company culture is top-down based is supported by the analysis results.

### "Betweenness" Centrality and mediating roles

While one stakeholder might have more direct connections to other stakeholders than another it is not the only thing that counts. The stakeholder who has a position between two stakeholders is called a mediator. He/she is passing the message from one stakeholder to the other. A mediator's position is strong because without him/her acting in between the two stakeholders, these couldn't communicate with each other. According to Krebs (2006), a mediator actor, which has high "Betweenness", has great influence on what flows within the network. Betweenness Centrality is calculated by looking at all the paths between all the stakeholders and calculating how many of those go through the specific stakeholder in question (Wasserman and Faust, 1994); a Relative value can be determined by dividing the previous value by the number of stakeholders in the network minus one.

When first analyzing Project 1 it can be seen that the Project Manager has clearly the highest Relative Betweenness Centrality value (41), indicating that he/she is the central mediator in the project. The Key Users are surprisingly on the second place in the list, but when looking at the aggregated graph from Project 1 the reason for their high position in the statistics gets clear: they are acting between the rest of the project organization and the end Users. In Project 2 the Business Process Owner gets the highest Relative Betweenness Centrality value of 100,5 , followed by the Program Manager with 29,1; the first high value is explained by the role as a mediator between the business process stakeholders and the project organization. When analyzing the Project 3 Betweenness values it can be seen that no stakeholder has as high values as in the other two projects: the Solution Owner has the highest Relative Betweenness Centrality value of 38,1 followed by the Business Deployment Manager (29,3), Business Owner (27), Project Manager (26,4), Process Owner (25,8) and the Program Manager (22,6). What is interesting in the results of Project 3 is that almost all stakeholders have some kind of mediator position. The results indicate that this project organization is more resistant to failures in the communication than in the other projects.

*Bottleneck roles*

A bottleneck is a process or stakeholder in any part of the organization that limits the throughput of the whole process (QMI Solutions, 2007). In an ERP project a potential bottleneck is a person who is also acting as a mediator between two parties. The Relative Betweenness Centrality values from the SNA-analysis maximized across projects shown in Table 2 give a good indication whether a person could be a potential bottleneck or not; are selected the stakeholder roles with the highest Relative Betweenness Centrality values.

| # | Stakeholder | Betweenn | Project |
|---|---|---|---|
| 1 | Business Process Owner | 100,519 | Project 2 |
| 2 | Project Manager | 41 | Project 1 |
| 3 | Solution Owner | 38,132 | Project 3 |
| 4 | Business Deployment Mgr | 29,337 | Project 3 |
| 5 | Program Mgr | 29,14 | Project 2 |
| 6 | Business Owner 1 | 27,027 | Project 3 |
| 7 | Global Rollout Mgr | 16,643 | Project 2 |
| 8 | Local Business Specialists | 16,583 | Project 2 |
| 9 | Global Project Mgr | 15,393 | Project 2 |
| 10 | Integration Manager | 12,792 | Project 3 |

Table 2: Highest Relative Betweenness Centrality Results

The most interesting detail is that the Business Process Owner, Project Manager and the Deployment Manager are all in the top of both the SNA analysis and the qualitative critical stakeholder list built by the interviewees. The SNA analysis results were not shown to the interviewees. This is why it can be said with a fairly high reliability that the Business Process Owner has the highest probability of becoming a bottleneck, followed by the Project Manager and the Local Deployment Manager. Three out of the top four bottlenecks roles are caused by a person at a managerial level.

*Conflicts*

As discussed in the literature review all stakeholders interpret the value of the project in their own way. The owners and users typically want to have as wide a scope as possible, the project staff wants to define the scope as accurately as possible and to freeze the design as early as possible, the finance managers look at the costs and the return on investment, and the top management looks at it from the company strategy perspective (Walsham, 1993). The only way to solve these conflicting objectives is to negotiate with all stakeholder groups until a consensus is reached (Remenyi, 1999).

Table 3 shows the data for the identified conflicts between the stakeholders, aggregated across projects and roles. The first three conflicts are internal conflicts within the project organization, while the Conflicts numbered from 4 to 10 are mainly between the business organization and the project organization. Conflict 9 incurs mainly between the Business Owners. The first column states the stakeholders who are involved in the conflict, the second column describes the conflict, followed by information on in which of the three interviewed projects the conflict was identified, the possible consequences of the conflict and finally recommendations on how to mitigate the risk of the conflict.

Most of the conflicts are related to how well the redesigned business processes and the new system are able to meet the business's requirements. Since one of the main objectives of a company-level ERP project is, in most cases, to harmonize the business processes within the company it means that many Business Units and especially the local business organizations need to change the way they are working currently. This causes natural resistance in the Business Units and local business organizations.

As in any complicated projects where there are many stakeholders involved, the importance of clear communication procedures, roles and channels are highlighted. Whether the communication happens inside the project organization or between the business and the project organizations, in all cases clear rules, roles and procedures must be set up. In most of the Conflicts (e.g. numbers 2, 4, 5, 6, 8, 9) the root reason for the conflict is that either parties (or both) feel that their needs and opinions are not taken into account well enough. It can be the case that some requests are not taken into account because they seem difficult to accommodate, but it might also happen that the other party doesn't know the requirements well or the argumentation behind them. Clear and fluent communication between the parties would enable to solve the conflict faster. If the problem escalates and the parties are not able to communicate well with each other, there should be a mediator between them. Quite often the Program or the Project Manager has to act as a mediator between two arguing parties.

| Stakeholders Involved | Conflict Description | Possible Consequences | Mitigation plan (How to control risk) |
|---|---|---|---|
| **Conflict 1:**<br>- Solution/Concept Owner (Process 1)<br>- Solution/Concept Owner (Process 2)<br>- Process Owner (Process 1)<br>- Process Owner (Process 2)<br>- Project Manager | **Solution specifications:** Solution Owners disagree about specific cross-process solutions in the ERP system. The goals of Finance and the Sales process Solution Owners could differ from each other in some cross module areas causing arguments between them. Project Manager has to mediate between the Solution Owners. The conflict might escalate up to the Process Owners. | Negative impact on schedule | Improve communication between teams |
| **Conflict 2:**<br>- Solution Owners<br>- Development Team<br>- Project Manager | **Solution specifications level of detail:** Arguments of the level of detail of the solution specifications. Development team finds the level too high and the Solution Owners believes it is detailed enough | Negative impact on schedule | Improve communication between teams |
| **Conflict 3:**<br>- Project Manager<br>- Project Members from the internal organization<br>- External Consultants | **Skills do not meet expectations:** The external consultants' skills and experience do not meet the Project Manager's expectations. The reason for the conflict is that the external consultancy company sells the resources to the client even though knowing that they are not skilled enough. | Negative impact on schedule (and quality) | Ensure the skills level before the contract (if possible) |
| **Conflict 4:**<br>- Business Owners<br>- Local Business Managers<br>- Program Manager | **Forming Program Steering Group:** Issues in forming the Program Steering Group. Some business organizations (e.g. Local Business Managers) see that they are not represented in the Steering Group. Arguments between the Business Owners and the Local Business Managers. Arguments between Business Owners. | Negative impact on businesses' commitment | Improve communication between business units/managers about the new to-be program objectives, scope, schedule |
| **Conflict 5:**<br>- Program Manager<br>- Project Manager<br>- Business Owner/Business Organization/Process Owner | **Resource Allocation:** Arguments about project resource allocation. Project needs resources from the Business Organization. Business Owner doesn't see the benefit of allocating the resources, which causes a conflict between the stakeholders | Negative impact on schedule and quality | Communicate to business the impact if needs not met. Escalate the issue as early as possible to Steering Group |
| **Conflict 6:**<br>- Local Steering Group/Local Country Management<br>- Global Deployment Manager<br>- Local Deployment Manager<br>- Program Manager<br>- Program Steering Group | **Project Scope or Solution Design:** Local Country Management wants changes in the solution. Local Deployment Manager promises changes to please the local management. The changes are not in line with the global project objectives, which causes a conflict between the program management (Global Deployment Manager, Program Manager, Program Steering Group) and the Local Country Management (Local Steering Group, Country Manager) and the Local Deployment Manager. | Negative impact on businesses' commitment | State project organizations roles and responsibilities clearly. Improve communication procedures to avoid escalation |
| **Conflict 7:**<br>- Local stakeholders (Local Steering Group, Business Managers, Business Process Specialists | **Changing Requirements:** Local Business Organization (Local Steering Group, Business Managers, and Business Process Specialists) changes their requirements after the deadlines for changes have been met. This causes a conflict between the project teams who are involved with designing and building the system (Dev. Team, Solution Owner, Technical Team) and the Local Business Organization. If the conflict escalates the Program Management and the Steering Group have to react on it. Even the Process Owner might get involved. | Negative impact on schedule | Communicate the deadlines better to |
| **Conflict 8:**<br>- Project<br>- Local Steering Group/Local Country Management<br>- Business Owners<br>- Global Deployment Manager<br>- Local Deployment Manager<br>- Program Manager<br>- Program Steering Group<br>- Process Owner | **Business Processes Change Resistance:** The Program Management wants to streamline the global business processes so that one solution would fit all. The local business organization resists the changes and doesn't see the project bringing any benefits for them (which might be the case). Business Owner does not support the changes and the project. The Program Steering Group has to interfere and possible escalate the issue even higher (CEO) to get the local business organization and the Business Owner to support the projects | | |
| **Conflict 9:**<br>- Business Owner from Business Unit 1<br>- Business Owner from Business Unit 2<br>- Program Steering Group | **Business Processes Disagreements:** The Business Units' (Business Owners) argue about the business processes, budget etc. due to different business needs. The Program Steering Group is involved in the discussion, since most Business Owner are members of the Steering Group. | | |
| **Conflict 10:**<br>- Local Deployment Manager/Team<br>- Key Users/Users | **User Resistance of Change:** Key Users or Users resistance to change because of fear of loosing power. | | |

Table 3: Identified conflicts amongst internal ERP stakeholders

When aggregating the conflicts into a matrix with impact factors, the highest risk for an ERP project is caused by resistance from the Business Owners (Conflict no. 8) and therefore the local business organization. The next highest risk is the resource allocation Conflict no. 5 which includes a conflict between the Business Owners.

*Relational Ties*

The social linkages between the actors are called *Relational ties*. The most common Relational tie categories relevant to this research are: formal relations (authority) and behavioral interaction (information sharing), in sum whether the relation is based on authority or a need to get or give information to another stakeholder or both. A matrix was built, based on the interview data, summarizing the relations between the stakeholders and explaining which type of relational tie exists between the stakeholders. Each column represents the stakeholder's (first row on the X-axis) relationship with the stakeholders on the Y-axis and the binary type of relational tie. "A" stands for official authority, meaning that the stakeholder has an authority over the other stakeholder; "B" is a relation which is based on information sharing, neither having official authority over each other. The Program Manager was found to have the most relations with official authority ("A"), followed by the Project Manager. The lower the stakeholder is situated in the organizational hierarchy, the more the relations were found to be based on information sharing ("B"). The Deployment Manager and the Deployment Team have the most relations which are not based on authority but on changing information. This is a logical result since they have to act with both the project organization and the local business organization stakeholders.

Another way of looking at the real authority in the network is to look at the strength of the relationships identified by the interviewees. The matrix mentioned before can be populated with the strength of influence the stakeholders have over each other (as collected in Figure 4), yielding the Relationship Strength matrix. It was found that the official authority ("A") and strength of the relation are strongly tied together. However it must be noticed that there are certain stakeholders who have, most likely because of specialist position, a high relation strength value, for instance the Development and Training teams. The Deployment Team's high value is explained by the fact that it has the most influence on the Key Users and the Users in the local organizations. Besides, the results have indicated that an ERP project has relatively equal relations based on information sharing ("B") between the internal stakeholders.

*Cliques*

A clique is a group of actors in which all actors have relations to all other actors. A clique has always a minimum of three nodes. By studying cliques we can identify tight groups within the ERP project internal stakeholder network. Some argue that the definition of a clique is too strict since it requires that everyone within a clique has a relation with everyone else, which is not always the case in real life environments (Wasserman and Faust, 1994). However, it still gives some picture of which subgroups exist within the network and that is why cliques were identified in this research.

When analyzing the Project sociomatrices with UCINET, or using the actor-by-actor clique co-membership matrix, in total 6 cliques were found in Project 1, 16 in Project 2, and 20 in Project 3, corresponding to increasing organizational spread.

*Generic Internal Stakeholder Network Model*

After having identified the most common internal ERP project stakeholders, the relations between them and the strength of their relations, a generic Internal Stakeholder Network model was built. The graph in Figure 5 below is a result of the entering the Relationship Strength matrix in NetDraw (Analytichtech NetDraw, 2006) software. It shows the key internal ERP project stakeholder roles, the relations between them and the strength of these relations in both directions.

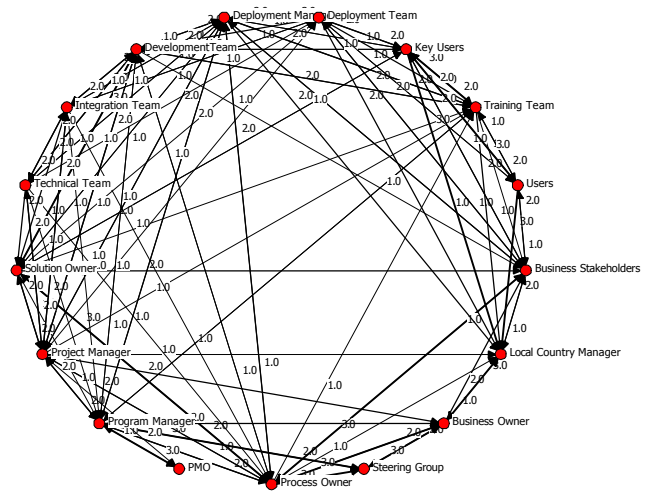

Figure 5: Generic Internal ERP project Stakeholder Network Model; the strength of influence is measured on a scale from 0 to 3, 0 being no influence, 1 being some influence, 2 strong influence and 3 very strong influence.

This model is based on the interview data. A few relations that were not mentioned by the internal interviewees but were seen as highly likely to have appeared, have been added to the data, but only after confirming the possibility of the relation with the external interviewees. The interview data included strength data for most of the relations. Since the interviewees did not determine a numerical value for the Steering Group's relation strengths the value was given by the researchers based on the external interviewees' comments about the Steering Group.

### *Stakeholder Drivers*

According to Herzberg (1959), responsibility, advancement, work itself and recognition for achievement are important for every employee because they indicate how competent a person is. The positive drivers were found to be in line with most IS projects. The negative drivers are the reasons that make the stakeholder resist something, in this case the ERP project. The negative drivers are found to be often fear-related, meaning that the stakeholder is afraid that the project's outcome is against his own best interest, such as decreasing his power in the organization (see the Survey section). It was observed that the stakeholders who were against the projects came in most cases from the business organizations, and not from within the project organizations.

### *Stakeholder Criticality per Project Phase*

One of the objectives of this research was to discuss the internal stakeholders influence on each project phase (Feasibility Study, Planning, Analysis, Design, Build, Test, Deploy and Hand-over to the support organization (Westland, 2006). In the Feasibility Study phase very few internal project stakeholders are involved except company management (CEO, CFO, CIO etc.) and the high level Business Unit management; in that phase the company management is the critical stakeholder group. More people join the project when the planning phase starts; at this stage the Program Manager, the Process Owners and the Business Owners are critical stakeholders. When the project evolves more stakeholders get involved with it: the Project Managers, Solution Owners and the Development Team become important. The Training Team and the Key Users' importance increases when the system is about to be implemented in the local business organizations.

## RECOMMENDATIONS FOR FUTURE ERP PROJECTS

Based on the experiences of past ERP projects, such as the three case studied in this research there are several recommendations that are suggested for future ERP project implementations. Some of them are generic project management advice and others suggestions about tools that could be used to mitigate some of the risks related to the internal stakeholders and their relations.

Business Process Redesign

- Ensure sufficient time for business process design and clarify the importance of proper design to the company management. The ERP system is built based on the business processes, which means that if the processes are not designed properly the system will not work properly either causing the project to miss its objectives.

Internal Stakeholder Motives

- Develop efficient and broad incentive programs for internal project stakeholders to ensure everyone's full commitment to the project and reduce power confrontations.

Project Progress

- Develop electronic real-time tools to track the progress of each stakeholder group in order to address issues on time. Such tools should include the real time analysis by SNA of email, Mobile email, SMS and Instant messaging traffic (not content); this real time capability would e.g. immediately spot bottleneck roles over time. Since the teams are dependent on each other's input/output, the progress tracking is extremely important.

Communication, Bottleneck and Conflict Resolution:

- Increase the number of stakeholder relations and the communication flow by encouraging the use of messaging tools within the project's stakeholder network to enable the stakeholders to communicate better with each other. This improves the projects "Closeness Centrality", creates more relations between the stakeholders, bringing them "closer" to each other. The risk of communication bottlenecks and conflicts is reduced.
- If allowed/possible, track the usage of the instant messaging tools and emails to gather information on the stakeholder relations and Relation Strength
- Analyze the usage of the project's shared document libraries, issue resolution tools etc. to gather knowledge of the activity level of the stakeholders and stakeholder cliques.
- Use SNA tools to visualize the stakeholder network and the real-time relations. Visualization helps to reveal possible bottlenecks or

weak points (mediators) in the organization.
- React immediately if there is lack of communication flow and density in the project
- Develop an efficient Governance Model, which defines the roles and responsibilities together with appropriate communication and issue resolution channels
- Organize common events to "weld" the stakeholders better together.

Role Staffing
- Ensure that the most critical roles identified in this research are staffed with people who have the correct skills and experiences.

CONCLUSION

The main conclusions, which can be utilized when planning new ERP projects, are the following:

1. ERP project's general stakeholders can be categorized in three groups: Project Organization, Business Organization and Company Management stakeholders. The Project Organization consists of the internal stakeholders who work full-time for the project. The Business organization consists of people who are influenced by the business process redesign and the new ERP system, and some are also internal to the project. Company Management consists of stakeholders who have stakes in the ERP project outcomes at business and/or operational level, such as the persons who decided to launch the project in the first place.

2. According to the interviews, supported by the SNA analysis, the internal stakeholders who are the most critical for an ERP project outcome are the company management Business Owner, Process Owners, the Deployment Manager, and the Solution Owner.

3. The internal stakeholders can be further divided into those who are responsible of the higher level planning and coordinating (Program Manager, Business Owner, Process Owner) and those who implement the plans in practice (Project Manager, Solution Owner, Development Team and Deployment Manager).

4. Next to the different criticality measures determined from the

interview data, the Social Network Analysis provided another interesting metric for analyzing the criticality of stakeholders in their network, the Closeness Centrality. The quantitative SNA results suggest that the Process Owner, Project Manager, Deployment Manager and Solution Owner are the internal stakeholders who have the most central roles in the project meaning that they have short paths to other stakeholders. These stakeholders should be utilized for communication and conflict solving purposes since they have the shortest and fastest connections to others. When considering communication it is important to have direct connections to the receivers to ensure timely and accurate messages. The same applies for conflict solving. The person who has direct connections to involved parties has higher chances of solving the conflict quickly.

5. Internal stakeholders who act as a bridge between other stakeholders, so called mediators, are extremely important actors in the network having a very strong position in the project. The Project Management should be aware of the risks related to the mediators. If the mediator fails to deliver the message, changes the message on the way, doesn't bring it on time or even at all, a bottleneck is formed. The following stakeholders were identified as mediators: Project Manager, Process Owner, Solution Owner, Deployment Manager and Business Owner

6. Project Management should understand that the conflicts that have the largest impact on the project's success derive from the relations between the project organization and the business organization or within the business organizations. The most serious conflicts that endanger the project are caused be business process redesign related issues. If the conflicts are not solved the dissatisfied stakeholders will not support the project and might even try to actively resist it.

7. To ensure everyone's commitment towards the project the personal objectives should be tied to the project's objectives. Each internal stakeholder thinks first "what is in it for me?" when they listen to project issues. If they don't gain anything they will most likely not commit to the project. The incentive plan should be made at least for the most critical internal business organization stakeholders such as the Business Owner.

Summarizing the recommendations on a general level, the Project Management should not ignore the sociological aspect of the ERP project implementation, and should not think that these aspects cannot be analyzed or monitored as shown in this paper. Too often the main focus of the Project Management is on the system and its design. Each project is slightly different but there are many similarities between them. If the above recommendations are followed some of the major project risks can be, if not fully avoided, at least minimized.

The above research results, and the generic internal ERP stakeholder network characteristics studied here, are also very useful for the progress in parallel research on business genetics which models the different forces whereby business units collaborate on an ad-hoc basis inside smart business networks, with fast connects and disconnects relying on ERP systems (Pau, 2006; Pau, 2008). Furthermore, one of the authors has developed specifically for ERP projects an add-on software enabling the real time social network dynamics to be analyzed while respecting work assignments and processes.